\documentclass[aps,prd,showpacs,nofootinbib,floats,floatfix,preprintnumbers,groupedaddress,twocolumn]{revtex4}

\usepackage{bm}
\usepackage{latexsym}
\usepackage{dcolumn}
\usepackage{amsmath,amsfonts,amssymb}
\usepackage{graphicx,epsfig}
\usepackage{amsmath}
\usepackage{fancyhdr}
\usepackage{hyperref}
\usepackage{graphicx,epstopdf}
\hypersetup{
	colorlinks   = true, 
	urlcolor     = blue, 
	linkcolor    = blue, 
	citecolor   = red 
}

\def\p{\partial}

\begin{document}
\title{Entropy corresponding to the interior of a Schwarzschild black hole}
\author{Bibhas Ranjan Majhi$^{a}$\footnote {\color{blue} bibhas.majhi@iitg.ernet.in}}
\author{Saurav Samanta$^{b,c}$\footnote {\color{blue} srvsmnt@gmail.com}}

\affiliation{$^a$Department of Physics, Indian Institute of Technology Guwahati, Guwahati 781039, Assam, India\\
$^b$Department of Physics, Narasinha Dutt College, 129, Belilious Road, Howrah 711101, India\\
$^c$Department of Physics, Bajkul Milani Mahavidyalaya, P.O. - Kismat Bajkul, Dist. - Purba Medinipur, Pin - 721655, India
}

\date{\today}

\begin{abstract}
Interior volume within the horizon of a black hole is a non-trivial concept which turns out to be very important to explain several issues in the context of quantum nature of black hole. Here we show that the entropy, contained by the {\it maximum} interior volume for massless modes, is proportional to the Bekenstein-Hawking expression. The proportionality constant is less than unity implying the horizon bears maximum entropy than that by the interior. The derivation is very systematic and free of any ambiguity. To do so the precise value of the energy of the modes, living in the interior, is derived by constraint analysis. Finally, the implications of the result are discussed.   
\end{abstract}

\maketitle

{\section{Introduction}}
   The concept of ``volume of a black hole'' attracted a lot of attention in recent years. Several ways have been proposed to define such quantity starting from the interior volume to thermodynamic volume. The later one is related to the modified first law of black hole thermodynamics \cite{Cvetic:2010jb,Dolan:2011xt} and for this case, the systems behave as the van der Waals system \cite{Kubiznak:2012wp,Majhi:2016txt}. The concept of interior volume is related to the space acquired by the black hole horizon. It was first introduced by Parikh \cite{Parikh:2005qs}. He showed that a definition of volume can be given which is independent on how one slices spacetime.  This idea was further developed by others \cite{Grumiller:2005zk}--\cite{Finch:2012fq}.  Although these works provide important understanding, the underlying concept is very much technical and not easy to visualize. Recently, a new way of defining the interior volume of the Schwarzschild black hole has been proposed \cite{Christodoulou:2014yia} by Christodoulou and Rovelli (from now on we call it as CR volume). For Kerr black hole, this calculation was done in  \cite{Bengtsson:2015zda} (For other types of black holes, see a very recent paper \cite{Wang}). Interestingly the interior volume increases linearly with the forward time and hence can be a candidate to resolve the information paradox problem. Discussions in this direction can be found in \cite{Ong:2015tua}--\cite{Ong:2016xcq}. Moreover, this volume is non-zero for an extremal black hole \cite{Bhaumik:2016sav}.

   Since the CR volume increases with time, at the end stage of the evaporation process, a black hole must have a large amount of volume to hide the huge information. Therefore one of the important quantities to be measured in this context is the entropy of the hidden modes within the CR volume. This is precisely the aim of the present paper. In this regard, let us mention that similar study has already been done in \cite{Zhang:2015gda} (also see \cite{Zhang:2016sjy}). Let us here point out the following inputs which were taken in that paper \cite{Zhang:2015gda} without not so clear justification. (i) In expanding the Klein-Gordon equation under the time dependent background, the scalar field was chosen as $\exp[-iET]\exp[iI(\lambda,\theta,\phi)]$ where $E$ is identified as the energy. But since the metric is time dependent it is not obvious if one can take such ansatz as there is no timelike Killing vector. (ii) To obtain the entropy from the free energy, the derivative with respect to the inverse temperature was taken. But the problem is in taking the derivative, CR volume was assumed to be independent of temperature. This is not correct because CR volume is a function of mass of the black hole and for Schwarzschild black hole, temperature is inversely proportional to mass. The expression for temperature in terms of mass has been used at the end and hence the question arises: is the calculation really consistent? (iii) Finally, the explicit use of the flux relation (Hawking expression) has been used. As pointed out in the paper itself \cite{Zhang:2015gda}, this expression is valid only when the mass is greater than Planck mass. In fact in the final stage $dM/dv\sim M$ and then one can not obtain the entropy proportional to area.
   
   In this paper we tried to get rid of the above inputs. Interpreting the expression of the integral form of the interior volume, given in \cite{Christodoulou:2014yia}, as the action and correspondingly defining an effective metric we find the Hamiltonian of a particle moving in this background.    Here the Hamiltonian is obtained by systematic constraint analysis. Then using this, the Gibbs' free energy of a massless particle is obtained. Interestingly we find that, the entropy of these massless modes, contained within the {\it maximum} interior volume is proportional to the horizon entropy \cite{Bekenstein:1973ur,Hawking:1974sw}. The proportionality constant, similar to \cite{Zhang:2015gda}, is less than unity-- which implies that the horizon contains the maximum entropy of a black hole. The whole calculation is self sufficient except we consider that the interior modes are in equilibrium with the horizon so that the system of massless modes has temperature equal to Hawking expression. 
   This is justified because similar concept is used in writing the first law of black hole mechanics in the form of the usual first law of thermodynamics. In this case, the role of temperature is played by the Hawking expression (the value measured by an infinitely distant observer); whereas the entropy is calculated on the horizon.

   We organize the paper as follows. In the next section a brief review of the results, given in \cite{Christodoulou:2014yia}, are mentioned. Mainly the effective metric for the interior volume, which is needed for our main purpose, is given. The energy for a particle, confined within this volume, is explicitly calculated in section \ref{Ham}. Using this, the entropy is calculated in the next one and then finally we conclude in section \ref{Con}.
   
\vskip 1mm
\section{\label{setup}Setup: a brief review}
The Schwarzschild metric in Eddington-Finkelstein coordinates takes the following form:
\begin{equation}
ds^2 = -f(r)dv^2+2dvdr+r^2d\Omega^2~, 
\end{equation}
where $f(r)=1-2M/r$ and $r=2M$ is the horizon. It has been shown in \cite{Christodoulou:2014yia} that the interior volume inside the horizon can be constructed by the surface $\Sigma\equiv\gamma\times S^2$ on which the induced metric is given by
\begin{equation}
ds_\Sigma^2 = \Big(-f\dot{v}^2+2\dot{v}\dot{r}\Big)d\lambda^2+r^2d\Omega^2~. 
\end{equation}
Here the curve is defined by $r=r(\lambda)$ and $v=v(\lambda)$, $\lambda$ being an arbitrary parameter. The volume turns out to be
\begin{equation}
V_{\Sigma} = 4\pi\int d\lambda\sqrt{r^4(-f\dot{v}^2+2\dot{v}\dot{r})}~.
\label{Volume}
\end{equation}
Note that, if we consider the integrant as a Lagrangian, then one can think it in terms of a metric
\begin{eqnarray}
ds^2_{\textrm{eff}} &=& r^4(-f\dot{v}^2+2\dot{v}\dot{r})d\lambda^2
\nonumber
\\
&=& r^4(-fdv^2+2dvdr)~.
\label{eff}
\end{eqnarray}
In the above Lagrangian, the coordinates are ($r,v$) and the corresponding momenta are ($P_r,P_v$). Therefore the phase-space volume is given by $\int drdvdP_rdP_v$. Our aim is to find the entropy contained within this phase-space volume. To calculate this quantity in a quantum statistical way, one needs to identify the Hamiltonian of a particle which is confined within this volume. This will be done in the following section. 

\section{\label{Ham}Hamiltonian}
Let us consider a particle of mass $m$ which is moving in a background metric
 \begin{eqnarray}
 ds_{\textrm{ansatz}}^2&=&g_{ab}dx^{a}dx^{b}
 \nonumber
 \\
 &=&-dt^2+r^4(-f(r)dv^2+2dvdr)~.
 \label{metric}
 \end{eqnarray}
 Here the Latin indices ($a,b,c$ etc.) run over all spacetime coordinates whereas the Greek indices ($\mu,\nu$ etc.) run over only space ones.
The metric is chosen in such a way that all the metric coefficients are independent of $t$ coordinate and hence there is a corresponding conserved quantity which is identified as the energy (Hamiltonian) of the particle. Moreover, the metric (\ref{eff}) for the interior volume corresponds to the $t=$ constant slice of the above manifold. Therefore it is obvious that the above one will serve our purpose. Below we find the Hamiltonian for this particle.  
 
    The action of a particle of mass $m$ moving in this space time is given by
 \begin{eqnarray}
S=m\int_1^2ds_{\textrm{ansatz}}=m\int_1^2(g_{ab}dx^{a}dx^{b})^{\frac{1}{2}}~.
 \end{eqnarray} 
 Above action has the reparametrization symmetry.
 The velocity of the particle is
 $u^{a}=(dx^{a}/{d\tau})$
 where $\tau$ is an arbitrary parameter by which the path of the particle is defined as $x^a=x^a(\tau)$. Introducing $\tau$ as an integration variable, action is written as
 \begin{eqnarray}
 S=m\int_1^2Ld\tau=m\int_1^2(g_{ab}\frac{dx^{a}}{d\tau}\frac{dx^{b}}{d\tau})^{\frac{1}{2}}d\tau~.
 \label{lagrangian}
 \end{eqnarray}
 So the Lagrangian is identified as
$L=m[g_{ab}(dx^{a}/{d\tau})(dx^{b}/{d\tau})]^{{1}/{2}}$.
Now the equations of motion of the system can be found easily from the Euler-Lagrange equations
 \begin{eqnarray}
 \frac{d}{d\tau}\left(\frac{\p L}{\p \dot{x^{a}}}\right)-\frac{\p L}{\p x^{a}}=0~.
 \end{eqnarray} 
 If $\tau$ is set as proper time (which is the affine parameter here), above equation gives the well known geodesic equation
  \begin{eqnarray}
 \frac{du^{a}}{d\tau}+\Gamma^{a}_{ \ bc}u^{b}u^{c}=0
 \label{geodesic}
 \end{eqnarray} 
 where $\Gamma^{a}_{ \ bc}$ is the Christoffel symbol which is symmetric in lower two indices.
So far we have not used any particular form of metric. So the geodesic equation obtained above is true for any spacetime. For the particular metric (\ref{metric}) note that $\Gamma^{0}_{ \ bc}$ has vanishing value and so from (\ref{geodesic}) the time part of the geodesic equation is given by
  \begin{eqnarray}
\frac{du^0}{d\tau}=0~.
 \end{eqnarray} 

Now to identify the correct Hamiltonian which represents the whole dynamics of the system, let us first concentrate on the canonical one.
Canonical momenta of the system is
 \begin{eqnarray}
 P_{a}=\frac{\p L}{\p \dot{x^{a}}}=\frac{m^2}{L}g_{ab}\frac{dx^{b}}{d\tau}~. 
 \label{momentum}
 \end{eqnarray}
 So the canonical Hamiltonian in this case is
 \begin{eqnarray}
 H_c=P_{a}\frac{dx^{a}}{d\tau}-L=0~.
 \end{eqnarray}
This vanishing value of canonical Hamiltonian is a typical signature of reparametrization invariant theory. Such speciality is not due to the curved background; it also happens for Minkowski spacetime as well.
For our present purpose it will be instructive to make a Hamiltonian analysis of the system (\ref{metric}). For flat spacetime, the analysis is done for chronological gauge in \cite{Hanson} and for proper time gauge in \cite{Gitman}. For curved spacetime, as far as we know, this study has not been done. So the constraint analysis presented here has independent importance apart from calculation of entropy.

Note that all four momenta are not independent
\begin{eqnarray}
P^2=g^{ab}P_{a}P_{b}=m^2~.
\label{PP}
 \end{eqnarray} 
 So we have a primary constraint
 \begin{eqnarray}
 \Psi=P^2-m^2\approx 0~. 
 \label{pri}
 \end{eqnarray} 
 Now following Dirac's algorithm \cite{Dirac}, total Hamiltonian is proportional to the primary constraint (\ref{pri}), i.e.
 \begin{eqnarray}
 H_T=\xi(\tau)\Psi=\xi(\tau)(P^2-m^2)~,
 \label{H_T}
 \end{eqnarray}
 where $\xi(\tau)$ is the proportionality constant and depends on parameter $\tau$ only.
 From this, Hamilton's equations are found to be
\begin{eqnarray}
\dot{x^{a}}=\frac{dx^{a}}{d\tau}=u^{a}=\{x^{a},H_T\}=2\xi P^{a}~,
\label{1st}
\end{eqnarray} 
and
 \begin{eqnarray}
\dot{P_{a}}=\{P_{a},H_T\}=-\frac{\p H}{\p x^{a}}=-\xi\frac{\p g^{bc}}{\p x^{a}}P_{b}P_{c}~.
 \end{eqnarray} 
 Using the definition of momentum variable (\ref{momentum}), first Hamilton's equation (\ref{1st}) is used to fix the Lagrange multiplier $\xi$:
\begin{eqnarray}
u^{a}=2\xi P^{a}=2\xi \frac{m^2}{L}u^{a}~,
\end{eqnarray} 
which  yields
\begin{eqnarray}
\xi=\frac{L}{2m^2}~.
\label{xi}
\end{eqnarray}
Thus total Hamiltonian (\ref{H_T}) comes out to be
\begin{eqnarray}
H_T=\frac{L}{2m^2} (P^2-m^2)~.
 \end{eqnarray} 
 Since there is only one constraint (\ref{pri}), the system is first class and hence it has gauge freedom\cite{Dirac}. This gauge freedom can be removed by imposing some condition on the arbitrary parameter $\tau$. It will be useful to interpret $\tau$ as proper time. Thus in the subsequent analysis we impose the gauge fixing constraint (proper time gauge)
  \begin{eqnarray}
\Phi_2=\frac{P^0}{m}\tau-x^0\approx 0~.
 \end{eqnarray} 
This together with the primary constraint (\ref{pri})
 \begin{eqnarray}
\Phi_1=P^2-m^2\approx 0
 \end{eqnarray} 
 make the system second class.
 
 Time consistency of the gauge fixing constraint $\Phi_2$ gives
 \begin{eqnarray}
&& \dot{\Phi}_2=\frac{\partial \Phi_2}{\partial t}+\{\Phi_2,H_T\}=0
\nonumber
\\
&& {\textrm{or}}, \ \frac{P^0}{m}-\{x^0,H_T\}+\frac{\tau}{m}\{P^0,H_T\}=0~.
\label{gauge}
\end{eqnarray}
 It is easy to calculate the second two terms of the left hand side,
 \begin{eqnarray}
\{x^0,H_T\}=\{x^0,\xi P^2\}=2\xi g^{ab}\{x^0,P_{a}\}P_{b}=2\xi P^0\label{x^0H_T}
 \end{eqnarray} 
 and
  \begin{eqnarray}
\{P^0,H_T\}=\xi\left(2P^{a}\frac{\partial g^{0 b}}{\p x^{a}}P_{b}-g^{0a}\frac{\partial g^{ bc}}{\p x^{a}}P_{b}P_{c}\right)~.\label{P^0H_T}
 \end{eqnarray}
 For our particular metric (\ref{metric}), right hand side of (\ref{P^0H_T}) vanishes. Thus we get from (\ref{gauge}) and (\ref{x^0H_T})
 \begin{eqnarray}
\xi=\frac{1}{2m}~.
 \end{eqnarray}
 Comparing with (\ref{xi}), above equation yields $L=m$.  Remembering the form of the Lagrangian (\ref{lagrangian}), we see $ds=\sqrt{g_{ab}dx^{a}dx^{b}}=d\tau$. This ensures that we are working with proper time gauge. This confirms that $\tau$ is indeed proper time. 
 
 Now the equations of motion can be calculated from Hamilton's equations: 
  \begin{eqnarray}
\dot x^{a}=\frac{1}{2m}\{x^{a},P^2\}=\frac{P^{a}}{m}
\label{dotx}
 \end{eqnarray} 
 and
 \begin{eqnarray}
\dot P^{a}&=&\frac{1}{2m}\{P^{a},P^2\}
\nonumber
\\
&=&\frac{1}{2m}\left(2\frac{\p g^{ab}}{\p x^{c}}g^{cd}-\frac{\p g^{db}}{\p x^{c}}g^{ca}\right)P_{d}P_{b}~.
 \end{eqnarray}
Using the identity $\p_c g^{ab}=-g^{ad}g^{be}\p_cg_{de}$, above equation can be simplified to
\begin{eqnarray}
\dot P^{a}=-\frac{1}{m}\Gamma^{a}_{ \ bc}P^{b}P^{c}~.
\label{dotp}
 \end{eqnarray}
Finally, making use of (\ref{dotx}) in the above equation one gets back the geodesic equation (\ref{geodesic}).

 To find the effective Hamiltonian of the system we first calculate the constraint matrix. This is
 \begin{eqnarray}
 C_{AB}=\{\Phi_{A},\Phi_{B}\}=2P^0
 \left(\begin{matrix} 0 & 1\\-1 & 0
 \end{matrix}\right),
 \end{eqnarray}
where $(A,B)=(1,2)$. Inverse of the constraint matrix is
  \begin{eqnarray}
( C_{AB})^{-1}=\frac{1}{2P^0}
 \left(\begin{matrix} 0 & -1\\1 & 0
 \end{matrix}\right)~.
 \end{eqnarray}
 Knowing the inverse constraint matrix, we can calculate the Dirac bracket $ \{f,g\}^*$ between two dynamical variables
 \begin{eqnarray}
 \{f,g\}^*= && \{f,g\}-\sum_{AB}\{f,\Phi_A\}(C_{AB})^{-1}\{\Phi_B,g\}
 \nonumber
 \\
 = && \{f,g\}+\frac{1}{2P^0}(\{f,P^2\}\{\Phi_2,g\}
 \nonumber
 \\
 && -\{f,\Phi_2\}\{P^2,g\})~.
 \end{eqnarray}
 The equation of motion of a dynamical variable should now be obtained from the following relation \cite{Dirac}
  \begin{eqnarray}
\dot{f}=\{f,H\}^*~.
\label{Dirac}
 \end{eqnarray}
Above equation is meaningful only if we can find an effective Hamiltonian $H$ of the system. This $H$ should be chosen in such a way that basic equations of motion (\ref{dotx}) and (\ref{dotp}) can be generated from (\ref{Dirac}). In fact $H=P^0$ serves our purpose. This can be checked in the following way. Let us calculate $\dot{x}^a$ and $\dot{P}^a$ from the above one. These are given by
\begin{eqnarray}
\dot x^{a}=\{x^{a},P^0\}^*=g^{a 0}+\frac{1}{P^0}\left(P^{a}+\frac{\tau}{m}g^{a 0}\Gamma^0_{ \ bc}P^{b}P^{c}\right) \  \ 
\end{eqnarray}
and
\begin{eqnarray}
\dot P^{a}=&&\{P^{a},P^0\}^*\nonumber \\= && \frac{\p g^{ab}}{\p x^{c}}g^{0c}P_{b}
\nonumber
\\
&&-\frac{1}{P^0}\left(\Gamma^{a}_{ \ bc}P^{b}P^{c}+g^{0a}\Gamma^{0}_{ \ bc}P^{b}P^{c}\right)~.
 \end{eqnarray}
Since $a=0$ has been eliminated by the gauge fixing constraint, only space component of $a$ ($a=\mu$) gives the required equations of motion:
  \begin{eqnarray}
\dot{x}^{\mu}=\frac{P^{\mu}}{P^0}; \,\,\  \dot{P}^{\mu}=\frac{1}{P^0}\Gamma^{\mu}_{ \ ab}P^{a}P^{b}~.
\label{EOM}
 \end{eqnarray}
$P^0$ can be eliminated from the above two equations to get the desired geodesic equation.
 
 Now since $P^0=-P_0$ we write (\ref{PP}) as
  \begin{eqnarray}
  g^{ab}P_{a}P_{b}=-(P^0)^2+2r^{-4}P_rP_v+fr^{-4}P_r^2=m^2.
  \end{eqnarray}
So the Hamiltonian i.e. the energy of particle is given by
\begin{equation}
\epsilon = P^0 = \Big(\frac{fP_r^2}{r^4}+\frac{2P_rP_v}{r^4}-m^2\Big)^{1/2}~.
\label{energy1}
\end{equation}
This can also be realized in the following manner. We know that, for usual relativistic particle the energy is related to the time component of the four momentum. Therefore calculation of $P^0$ from (\ref{PP}) must lead to the above value. This intuitive understanding helps us to know the energy for a massless particle (like photon) when it moves in the background metric (\ref{metric}). This is found by taking $m\rightarrow 0$ limit in the above equation. So for photon
\begin{equation}
\epsilon =  \Big(\frac{fP_r^2}{r^4}+\frac{2P_rP_v}{r^4}\Big)^{1/2}~.
\label{energy}
\end{equation}
This expression of energy will be used in the following section.
\section{\label{main}Calculation of entropy}
Let us now calculate the entropy corresponding to the energy (\ref{energy}) for a massless particle. This will lead to the entropy of modes confined within the interior volume of the Schwarzschild black hole if we assume that the system contains only these modes. Since there is no restriction on its number, we do not have the chemical potential term and therefore the Gibbs' free energy is given by
\begin{equation}
G=-\frac{1}{\beta}\ln\mathcal{Z}= \frac{1}{\beta}\sum_{\epsilon}\ln\Big(1-e^{-\beta\epsilon}\Big)~,
\label{Gibbs}
\end{equation}
where $\beta$ is the inverse temperature and $\mathcal{Z}$ is the grand canonical partition function.

  Now for the present system we have the following conditions. To obtain the maximum volume of the interior of the horizon, imposition of extremality lead to the constant value of the radial coordinate; i.e. one must have $\dot{r}=0$. For the present Hamiltonian, the equation of motion for $r$ can be obtained from (\ref{EOM}), which leads to $\dot{r} = r^{-4}(P_v+fP_r)/P^0$. It gives one condition $P_v+fP_r=0$. Also it was considered that both radial and ingoing null coordinates are function of the parameter $\lambda$; i.e. $r=r(\lambda)$ and $v=v(\lambda)$. Therefore, in general we must have $v=F(r)$ where $F(r)$ is some function of the radial coordinate (See \cite{Christodoulou:2014yia} for details). Hence the Gibbs' free energy in this case should be calculated from
  \begin{equation}
  G = \frac{1}{\beta}\int \frac{dP_rdP_vdrdv}{h^2}\ln(1-e^{-\beta\epsilon})\delta(P_v+fP_r)\delta(v-F(r))~.
  \end{equation}
Now since $\epsilon$ is given by (\ref{energy}), the two integrations on $P_v$ and $v$ can be evaluated easily by the help of two Dirac delta functions. This gives
\begin{equation}
G=\frac{1}{h^2\beta}\int dP_r dr \ln\Big[1-e^{-\beta\Big(\frac{-fP_r^2}{r^4}\Big)^{1/2}}\Big]~. 
\end{equation}
Here the integration on $P_r$ runs from zero to infinity. Therefore using integration by parts we obtain
\begin{equation}
G=-\frac{1}{h^2}\int dr \frac{\sqrt{-f}}{r^2}\int_0^\infty dP_r \frac{P_r}{e^{\frac{\beta\sqrt{-f}P_r}{r^2}}-1}~.
\end{equation}
Since we are calculating the above quantity for interior volume, the radial coordinate runs from $2M$ to zero. Therefore defining $(\beta\sqrt{-f}/r^2)P_r$ as $x$ and $r/2M$ as $y$ one finds
\begin{equation}
G=-\frac{8M^3}{h^2\beta^2}XY~,
\label{G}
\end{equation}
where
\begin{eqnarray}
&&X=\int_0^{\infty}\frac{xdx}{e^x-1} = \frac{\pi^2}{6};
\nonumber
\\
&&Y=\int_1^0 \frac{y^{5/2}}{\sqrt{1-y}} = -\frac{5\pi}{16}~.
\label{XY}
\end{eqnarray}

  Now for Schwarzschild black hole, the inverse temperature is given by $\beta = 8\pi M/\hbar$ and so the Gibbs' free energy in this case turns out to be
  \begin{equation}
  G=-\frac{\hbar\beta}{32\pi^5}XY~. 
  \end{equation}
Therefore the entropy is found to be
\begin{equation}
S=\beta^2\frac{\partial G}{\partial\beta} = -\frac{\hbar}{32\pi^5}\beta^2XY.
\end{equation}
Next using the value of the horizon area $A=16\pi M^2$ and the numerical values of $X$ and $Y$ from (\ref{XY}), we find the entropy corresponding to the maximum interior volume as
\begin{equation}
 S = \frac{5}{192\pi}S_{BH}~.
 \label{entropy}
\end{equation}
where $S_{BH} = A/4\hbar$ is the entropy on the horizon of the black hole. It shows that the interior volume contains less entropy compared to that contained on the horizon.

Before concluding, let us mention about one important input in the above calculation. We considered that, the inside massless modes are at a temperature which is equal to that of the event horizon. The justification is the following. It must be remembered that the CR volume is the interior portion of the horizon whose one boundary is the horizon. This can be clarified from \cite{Christodoulou:2014yia}, as the radial coordinate runs from $r=2M$ to $r=0$ which has been used in the above calculation also (see the discussion above (\ref{G})). So our system is like a black body which encloses massless particles. This system is in equilibrium and its temperature is Hawking temperature. Therefore the particles are also at inverse temperature $\beta=8\pi M/\hbar$. However, due to Hawking radiation, the temperature of the horizon changes as the radius of the horizon is changing. But we assume that this process is slow enough (i.e. quasi-static) to equilibrate the whole system within the CR volume. This assumption is justified because the rate of mass loss due to Hawking radiation (for a Schwarzschild black hole) is $\frac{dM}{dv}\sim-\frac{1}{M^2}$. So except at the end state of evaporation, evaporation process is slow \cite{Ong:2015dja, Zhang:2015gda}. This is an important input as our calculation is valid only in the equilibrium situation. 

\section{\label{Con}Conclusions}
  CR volume is the maximum interior of the horizon which increases linearly with time \cite{Christodoulou:2014yia}. It is expected that, this can have enough space to hide the information at the end stage of black hole evaporation. Moreover, the extremal black holes also have non-vanishing CR volume \cite{Bhaumik:2016sav}. Therefore, such a non-trivial concept can be a candidate for resolving the information paradox. Since according to the information theory, these hidden information can be retrieved in terms of entropy, it would be interesting to calculate the entropy contained within this. The detailed discussion on the role of CR volume in information paradox can be followed from \cite{Ong:2015tua}. As we do not want to repeat the same we just mentioned the relevant literature.
  
  In this paper we precisely addressed the same. We calculated the entropy contained by the CR volume for the Schwarzschild black hole where there are only massless modes. The approach adopted here is statistical one. In doing so, we first interpreted the integrand of the interior volume expression as an effective metric. Then using this, the energy of the modes was identified. Since the canonical Hamiltonian vanishes, one had to use the method of constraint analysis to handle the situation. To the best of our knowledge, the proper calculation of the Hamiltonian is itself new. Finally using the energy, we calculated the entropy by evaluating the Gibbs' free energy through statistical mechanics. We think, this approach is free of some of the not so clear assumptions considered in \cite{Zhang:2015gda} (see the discussion in the second paragraph of the introduction).
  
  The outcome is very interesting. The entropy turns out to be proportional to the horizon entropy (Bekenstein-Hawking expression) and the proportionality constant is less than unity. This implies that, the horizon contains maximum entropy. Moreover, since we have a non-zero value, one can argue that the whole entropy of the black hole does not resides only on the horizon. Therefore it may be possible that such quantity plays a bigger role in the context of information paradox problem. Finally we want to mention that, our present calculation is much more robust than the earlier one \cite{Zhang:2015gda} and hence the result -- dependence of the CR entropy on the horizon area -- is now further bolstered. Therefore we hope that our calculation can give deeper insight into the precise role of CR volume in the quantum nature of black hole. But at this stage there is no concrete evidence and obviously further investigation is needed. 
  
   In this connection, it may be worth mentioning one recent work \cite{Kawai:2015uya} which also investigated the contribution to the entropy from the interior. But the setup is very different from the present one. There the authors analyze the time evolution of a spherically symmetric collapsing matter including the back reaction from the evaporation. The solution of the semi-classical Einstein equations indicates that the collapsing matter forms a dense object and evaporates without horizon or singularity. Using the obtained interior metric, the entropy density has been evaluated. Integrating it over the proper volume of the interior region one obtains the area law (also see \cite{Kawai:2017txu}). Hope all these will give some insight to the information paradox problem.

\vskip 4mm
{\section*{Acknowledgments}}
\noindent
We thank Debraj Roy and Sujoy Kumar Modak for several useful discussions. 
The research of one of the authors (BRM) is supported by a START-UP RESEARCH GRANT (No. SG/PHY/P/BRM/01) from Indian Institute of Technology
Guwahati, India. 

\end{document}